\documentclass[aps,twocolumn,amsmath,ams-fonts,superscriptaddress]{revtex4}
\usepackage[dvipdfmx]{graphicx}
\usepackage{dcolumn}
\usepackage{bm}
\usepackage{here}
\usepackage{relsize}
\usepackage{amssymb}
\usepackage{txfonts}
\usepackage{braket}
\usepackage{color}

\usepackage[pdftex]{hyperref}   
\def\[#1\]{\begin{align}#1\end{align}}

\begin{document}
\title{
Adiabatic and nonadiabatic spin torques induced by spin-triplet supercurrent
}
\author{Rina Takashima}
\email{takashima@scphys.kyoto-u.ac.jp}
\affiliation{Department of Physics, Kyoto University, Kyoto 606-8502, Japan}
\author{Satoshi Fujimoto}
\affiliation{Department of Materials Engineering Science, Osaka University, Toyonaka 560-8531, Japan}
\author{Takehito Yokoyama}
\affiliation{Department of Physics, Tokyo Institute of Technology, Tokyo 152-8551, Japan}
\date{\today}
\begin{abstract}
We study spin transfer torques induced by a spin-triplet supercurrent in a magnet with the superconducting proximity effect.     
By a perturbative approach, we show that spin-triplet correlations realize  new types of torques, which are analogous to the adiabatic and nonadiabatic ($\beta$) torques,  without extrinsic spin-flip scattering.  Remarkable advantages compared to conventional spin-transfer torques are highlighted in domain wall manipulation. 
Oscillatory motions of a domain wall do not occur for a small Gilbert damping, and the threshold current density to drive its motion becomes zero in the absence of extrinsic pinning potentials due to the nonadiabatic torque controlled by the triplet correlations.  
\end{abstract}
\maketitle 

Efficient manipulation of magnetization is of great technological importance. Spin-transfer torques (STT), which can control magnetization with an electric current, have attracted attention\cite{Slonczewski1996,Berger1996, Katine2000,Klaui2003, Grollier2004,Lim2004, Yamaguchi2004}, and STT can be applied to the so-called racetrack memory using magnetic domain walls\cite{Parkin2008}.
 In a smooth magnetic texture $\bm n$, 
spin-polarized currents exert STT on $\bm n$, which is  given by  
\cite{Haney2008,Tatara2008,Tserkovnyak2008}
\[\bm \tau_{\rm STT}=
- (\bm j_s \cdot \bm \nabla)  \bm n+ \beta \bm n \times (\bm j_s \cdot \bm \nabla) \bm n. \label{tor_0}
\]
Here $\bm j_s= -(Pa^3/2eS) \bm j$, and $\beta$ is a dimensionless parameter, where $\bm j$ is a charge current density,  $P$ is the spin polarization of current, $a$ is the lattice constant, $S$ is the spin size, and $-e$ is the electron charge. The first term in Eq.~\eqref{tor_0} arises when the electron spins follow the texture adiabatically.  The second term, often referred to as the non-adiabatic torque, is known to have two origins \cite{Zhang2004, Xiao2006,Tatara2008}. 
It appears from spin-flip impurity  scatterings or the spin-orbit coupling. 
 It also occurs when electrons fail to follow magnetic textures because the texture is not smooth enough. 
As demonstrated in several works \cite{Thiaville2005,Tatara2008}, the nonadiabatic torque plays a crucial role in magnetization dynamics. 
For $\beta\neq 0$, the threshold current density for a steady motion of a domain wall becomes zero in the absence of pinning potentials. 

Recently, superconductivity has opened up new possibilities for spintronics with suppressed Joule heating\cite{Linder2015, Eschrig2015}.  
It has been pointed out that the Josephson current exerts a spin torque on magnetization in ferromagnetic Josephson junctions\cite{Waintal2002, Tserkovnyak2002, Zhao2008, Braude2008, Konschelle2009,Linder2011,Linder2012}.  
{Furthermore, with spin valves using superconductors\cite{Buzdin1999,Tagirov1999, Leksin2012, Banerjee2014, Singh2015}, one can change the resistance drastically by a magnetic field}, and the lifetime of spin density is enhanced in a superconducting state relative to a normal state\cite{Yang2010, Quay2013, Hubler2012}.     
{Such an interplay of  superconductivity and magnetic moments is important especially with spin-triplet Cooper pairs due to the coupling between triplet order parameters and localized moments\cite{Kastening2006, Yokoyama2011, Brydon2008b, Brydon2009, Brydon2011, Yokoyama2015b}.} 
Triplet pairs can arise in the interface between a ferromagnet and singlet superconductor when there is   
 magnetic inhomogeneity\cite{Bergeret2005} or spin-orbit couplings
\cite{Bergeret2013, Bergeret2014}. Experimentally, the proximity effect of triplet pairs has been observed in fully-spin polarized metals\cite{Keizer2006, Anwar2010} and multilayers with noncollinear magnets\cite{Khaire2010, Robinson2010}. 
The spin-triplet proximity effect to a ferromagnet from Sr$_{2}$RuO$_{4}$, a candidate of a triplet superconductor, has been also observed\cite{Anwar2016}. 

Given the experimental advances in the proximity-induced triplet Cooper pairs in magnets, triplet supercurrent-induced STT is a promising way to realize an efficient control of magnetization. Utilization of a supercurrent suppresses Joule heating and the tunablity of STT may be enhanced by pairing degrees of freedom. 
However, while several works showed that a supercurrent exerts a spin-torque in ferromagnetic Josepshon junctions\cite{Waintal2002, Tserkovnyak2002, Zhao2008, Braude2008, Konschelle2009,Linder2011, Linder2012}, 
it still remains unclear how a triplet supercurrent acts on a localized moment, and how STT is changed by triplet-paring correlation.  


In this work, we microscopically study STT induced by triplet supercurrents considering  the spin-triplet proximity effect. 
We show that the derived STT have two parts, analogous to the adiabatic and non-adiabatic torques,  which can be tuned by the triplet correlations.
 Remarkable advantages compared to conventional STT are highlighted in domain wall manipulation. 
 In contrast to the non-adiabatic STT in normal metals, the supercurrent-induced STT do not require extrinsic scattering processes, and hence, is more easy to  control. Furthermore, a domain wall does not show oscillatory motions for a small Gilbert damping, and hence an efficient manipulation can be realized. 
 
Let us consider a thin-film magnet with the proximity effect of  $p$-wave triplet superconductivity modeled by the Hamiltonian :
\[
H_{\rm el} =&
-t \sum_{\langle i,j \rangle }c^{\dag}_{i \alpha}  c_{j\alpha} 
-\mu \sum_i c^{\dag}_{i \alpha} c_{i\alpha}\nonumber\\
&-J_{\rm sd} S\sum_i  \bm n(\bm r_i) \cdot \bm \sigma_{\alpha \beta}c^{\dag}_{i \alpha} c_{i \beta} \nonumber\\
&+\frac{\Delta_0}{2}\sum_{i, j} e^{i \bm Q \cdot (\bm r_i +\bm r_j) } \left[ (\bm{d}_{ ij}  \cdot \bm \sigma)i \sigma^y\right]_{\alpha \beta}
 c^{\dag}_{i \alpha} c^{\dag}_{j \beta}+{\rm H.c.} \label{ham}\]
Here $c^\dag_{i\alpha}(c_{i\alpha} ) $  is the electron creation (annihilation) operator at site $i$ with spin $\alpha$ on a square lattice, $\langle i, j \rangle $ is taken over the nearest neighbor pairs, $t$ is the hopping amplitude, $\mu$ is the chemical potential, and $\bm \sigma =(\sigma^x, \sigma^y,\sigma^z)$ are the Pauli matrices. Electrons couple to localized spins given by $S\bm n(\bm r_i)=S(\sin \theta \cos \phi, \sin \theta \cos \phi,\cos \theta) $ with the coupling constant $J_{\rm sd}$. The spin texture $\bm n(\bm r)$ varies smoothly with the length scale $\ell\  (\gg \xi_{\rm SC})$ where {$\xi_{\rm SC}$} denotes the superconducting coherence length.   
The last term in Eq.~\eqref{ham} is the proximity-induced triplet $p$-wave pairing, given by $\bm d_{ ij}=(d^x_{ ij},d^y_{ ij},d^z_{ ij})$, where $\bm d_{ij} =- \bm d_{ji}$, and $\Delta_0$  is the pairing amplitude.

\begin{figure}
\includegraphics[width=6.cm]{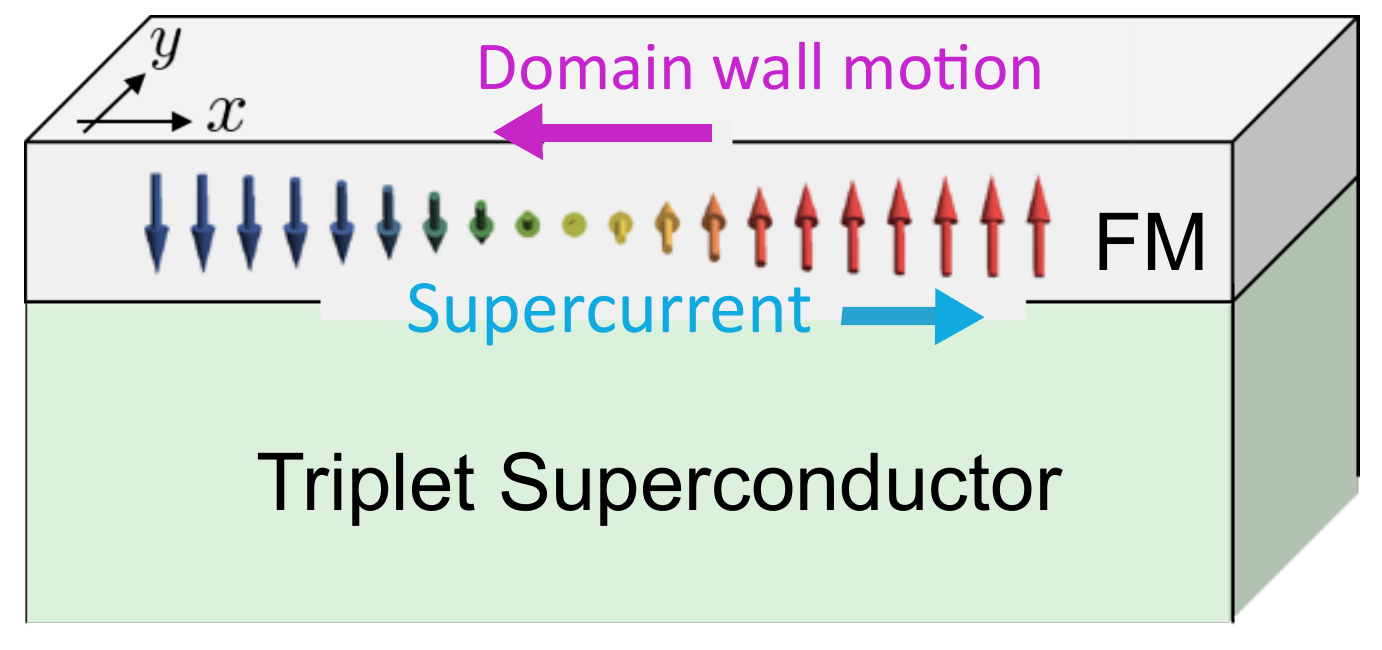}
\caption{ (Color online)
Proposed setup. A ferromagnet (FM) with a domain wall is attached on a superconductor, which produces a spin-triplet proximity effect. 
 We apply a supercurrent in the $xy $ plane and drive the domain wall motion. 
}
\label{setup}
\end{figure}

The phase gradient of the paring function describes a supercurrent.   
We consider a phase given by $ \bm {Q}  \cdot (\bm r_i +\bm r_j)$, which results in  the supercurrent density $\bm j \simeq  -2 t  e n_e a^2  \bm Q  $ (for  $|\bm Q| a \ll 1$ and at low temperature), where $n_e $  is  an  electron density\footnote{For a high density, $n_e$ should be replaced with an effective electron density modified by lattice effects. } that participates in the supercurrent. Here, the supercurrent can be supplied from an external dc current source.
We also note that we can restore the gauge invariance by redefining $\bm Q$ in the current so as to include the vector potential.   
As shown in Fig.~\ref{setup}, a relevant experimental setup of the above model is  a heterostructure composed of a metallic magnet and a triplet superconductor  (e.g., Sr$_2$RuO$_4$).  
A triplet superconductor can be replaced by a singlet superconductor (e.g., Nb) with a conical magnetic layer such as Ho \cite{Robinson2010} or the spin-orbit coupling\cite{Bergeret2013, Bergeret2014}, which produces the triplet proximity effect. 

To derive STT, we calculate the spin density induced by a supercurrent to the linear order of $\bm j  \propto \bm Q $. 
In the following calculation, we perturbatively treat the spatial derivative of $\bm n(\bm r_i) $.   
This treatment can be simplified by rewriting the Hamiltonian with electron operators $ a_{i \alpha} $, the spin quantization axis of which is parallel to $\bm n(\bm r_i) $\cite{Tatara2008}. It is defined by $c_{i \alpha}=(U(\bm r_{i}))_{ \alpha \beta} a_{i \beta}$, where $U(\bm r_i)= \bm m(\bm r_i) \cdot \bm \sigma $ is a unitary matrix, and  $\bm m(\bm r_i) =(\sin \left(\theta/2\right) \cos \phi, \sin \left(\theta/2\right) \sin \phi,\cos \left(\theta/2\right) ) $ with  $\theta$ and $\phi$ being the angles of $\bm n(\bm r_i)$.  This satisfies $U^\dag (\bm n \cdot \bm \sigma)U=\sigma^z$. 
The Hamiltonian Eq.~\eqref{ham} is rewritten as 
\[
H_{\rm el}\simeq& 
 \sum_{\bm k}( \xi_{\bm k} \mathbb I_2 -J_{\rm sd} S \sigma^z  )_{\alpha \alpha} a^{\dag}_{\bm k \alpha}  a_{\bm k\alpha}\nonumber \\
&+\sum_{\bm k, \bm q} 
\varv^\nu_{\bm k +\bm q/2} A^a_{\nu}(\bm q) \sigma^a_{\alpha \beta}a^\dag_{\bm k+\bm q \alpha}  a_{\bm k \beta}
\nonumber\\
&+\frac{1}{2}\sum_{\bm k} 
\Delta^{\alpha }_{\bm k}
a_{\bm k+\bm Q \alpha}^\dag  a_{-\bm k+\bm Q \alpha}^\dag+\text{H.c. },
\label{ham2} \]
where {
 $\xi_{\bm k}=-2t (\cos (k_x a)+\cos (k_y a)) -\mu $ is the kinetic energy, $\varv_{\bm k}^\nu =\partial \xi_{\bm k}/ \partial k_\nu $ is the velocity,} $\mathbb I_2 = {\rm diag } (1,1)$ , and $a_{\bm k \alpha}$ is the Fourier transform of $a_{i \alpha}$.
The second term arises from the hopping in the presence of a noncollinear texture. {Here we define the spin gauge field} $ A_{\nu i}^a \sigma^a=- i  U(\bm r_i)  \partial_\nu U(\bm r_i) $, where we denote $A_\nu^b  (\bm q )=N^{-1}\sum_i A_{\nu i }^b e^{-i \bm q \cdot \bm r_i}$ with {$\nu=\{x, y, z\}$ and $b=\{x, y, z\}$}.  $N$ is the total number of sites. 
Assuming a large exchange splitting $\sim 2J_{\rm sd}S \gg |\Delta_0|$, we focus on the equal spin pairing given by 
\[
\Delta^{\alpha}_{\bm k} 
&=\Delta_0 (U^\dag (\bm{d} (\bm{k} ) \cdot \bm \sigma)i \sigma^y U^* )_{\alpha \alpha}, \label{dvec}\\
&=-\Delta_0\left( (\mathcal R^{ab}d^b(\bm k)\sigma^a) i \sigma^y\right)_{\alpha \alpha},  \]
where $\bm d(\bm k) =  N^{-1}\sum_{\langle ij \rangle } \bm d_{ij}e^{-i \bm k \cdot (\bm r_i- \bm r_j)}$.  We neglect  pairings between spin-split bands, which correspond to the components of $\bm d(\bm k)$ parallel to $\bm n$.  
$\mathcal R^{ab}= 2m^a m^b -\delta^{ab}\  {(a, b =\{x, y, z\})}$  is a SO$(3)$ rotation matrix corresponding to the unitary matrix $U$.   
In the last term in Eq.~\eqref{ham2}, we neglect the spatial dependence of $\theta, \phi $, terms of the order of $\xi_{\rm SC}/\ell $,  assuming that $|\Delta_0(\bm k)| /t$ is also a small parameter. 

The spin expectation value of electrons $
 s^a_{ i}=\frac{1}{2}
( \sigma^a)_{\alpha \beta} \langle c^{\dag}_{i \alpha}   c_{i \beta} \rangle 
$  can be described by the operator $a_{i \alpha}$  as 
  \[
s^a_{i} =\mathcal R^{ab}(\bm r_i) \tilde s^b_{i }, \label{spin_c} 
\]
where  
$
\tilde s^a_i= \frac{1}{2}
( \sigma^a)_{\alpha \beta}  \langle a^{\dag}_{i \alpha}   a_{i \beta}\rangle
\label{spin1}$. 
Noting this relation, we obtain the spin density induced by a supercurrent as 
\[
\delta \tilde s^a_{\bm q}  &\coloneqq\left(
\lim_{ \bm Q \rightarrow \bm 0}
\frac{\tilde s^a_{\bm q} -\tilde s^a_{\bm q}|_{\bm Q= \bm 0}}{ Q_\eta }   
\right)\frac{ j_\eta}{(-2t en_e a^2)},\\
&=   \pi_{\nu \eta}^{ab} A^b_\nu(\bm q) \frac{ j_\eta}{2en_e }, \label{spin2}
\]
 where  $\delta \tilde{\bm s}_{\bm q} =N^{-1}\sum_i \delta \tilde{\bm s}_i e^{-i \bm q \cdot \bm r_i}$ and 
\[
&\pi_{\nu \eta}^{ab} = 
\lim_{\bm q\rightarrow \bm 0}
\frac{-T  }{4N ta^2} \sum_{n, \bm k}  \frac{\partial^2 \xi_{\bm k} }{\partial k_\nu \partial k_\eta}
{\rm Tr} [ S^a {\mathcal G_{\bm k+\bm q}(i {\epsilon_n})
S^{b}   \mathcal G_{\bm k}(i {\epsilon_n})}].
\label{chi}\]
See Supplemental Materials (SM) for detail \cite{SM}.
Here, 
$\mathcal G_{\bm k} (i \epsilon_n)$ is the Green function in Nambu representation, the basis of which is 
$(a_{\bm k \uparrow},a_{\bm k \downarrow}, a^\dag_{-\bm k \uparrow},a^\dag_{-\bm k \downarrow} )^T$, and
$S^a$ is a spin matrix in the Nambu representation, which is given by  
\[S^a &=
\begin{pmatrix}
\sigma^a & \\
 & -(\sigma^{a})^T
\end{pmatrix}.
\]  
Its inverse is defined by $(\mathcal G_{\bm k}(i \varepsilon_n) )^{-1}
=i \epsilon_n \mathbb I_4 -H_{\rm BdG}(\bm k)$, where
\[
H_{\rm BdG}(\bm k) &=
\begin{pmatrix}
\xi_{\bm k} \mathbb I_2 -M \sigma^z & \Delta(\bm k)  \\
\Delta^*(\bm k)  & -\xi_{\bm k} \mathbb I_2 +M \sigma^z 
 \end{pmatrix},
\]
 $\mathbb I_4={\rm diag}(1,1,1,1)$, $\epsilon_n = \pi {  T} (2 n+1) $ is Matsubara frequency with temperature $T$, and $\Delta(\bm k) =\rm{diag}(\Delta^{\uparrow}_{\bm k}, \Delta^{\downarrow}_{\bm k})$.

In Eq.~\eqref{chi},  we have neglected terms which are vanishingly small at low temperatures compared to the critical temperature, in a system with a full gap or point nodes, see SM \cite{SM}.  Also, we have taken the limit $\bm q\rightarrow \bm 0$ assuming that the momentum transfer from a smooth magnetic texture is small compared  to the Fermi momentum \cite{Tatara2008}. 
{We note that $\bm s_i$ is invariant by a unitary transformation $U \rightarrow Ue^{i \varphi \sigma^z} $ with an arbitrary spin rotational angle $\varphi(\bm r)$ around $\bm n$, 
while $\tilde{\bm s}_i$ and $\mathcal R^{ab} $ change their forms.   
In the following,  we explicitly use  $\frac{\partial^2 \xi_{\bm k} }{\partial k_\nu \partial k_\eta} \propto \delta_{\nu \eta}$ and define $\pi^{ab}_{\nu\eta}= \pi^{ab}_{\nu} \delta_{ \nu \eta}  $. 

From Eqs.~\eqref{spin_c} and \eqref{spin2}, 
we obtain the local STT, $\bm \tau_{\rm STT}=2 J_{\rm sd} \bm n \times \delta\bm s_i$,  as 
\[\bm \tau_{\rm STT}=\sum_{\nu=x, y}
\frac{-\tilde{P}_\nu a^3}{2eS} j_\nu \left( 
-\partial_\nu \bm n +\tilde \beta_\nu \bm n \times \partial_\nu \bm n \right).
\label{pol1}
\]
Here $\tilde P_\nu$ and $\tilde \beta _\nu $ are the analogs of the spin polarization of a current $P$ and  $\beta$ in Eq.~\eqref{tor_0}, and they are given by  
\begin{widetext}
\[
\tilde{P}_\nu &= \frac{J_{\rm sd} S}{  n_e a^3}
\left[ \frac{1}{2}\left(\pi^{xx}_{\nu} +\pi^{yy}_{\nu}\right) 
+\frac{1}{| \partial_\nu \bm n|^2}  \left( -\pi^{(1)}_{\nu} \left ((\partial_\nu \theta)^2 -\sin^2 \theta (\partial_\nu \phi)^2 \right) +2\pi^{(2)}_{\nu}\sin \theta \partial_\nu \theta \partial_\nu \phi \right)
\right],\\
\tilde{\beta}_\nu &=-\frac{J_{\rm sd} S}{  n_e a^3} \frac{1}{ \tilde{P}_\nu }
\frac{1}{| \partial_\nu \bm n|^2} \left( \pi^{(2)}_{\nu}\left ((\partial_\nu \theta)^2 -\sin^2 \theta (\partial_\nu \phi)^2 \right) +2\pi^{(1)}_{\nu}\sin \theta \partial_\nu \theta \partial_\nu \phi \right),\label{beta_t} 
\]
\end{widetext}
where 
\[
&\pi^{(1)}_{\nu} =\cos (2 \phi) \frac{1}{2}\left(\pi^{xx}_{\nu} -\pi^{yy}_{\nu}\right) +\sin (2 \phi) \pi^{xy}_{\nu},\\
&\pi^{(2)}_{\nu} =\sin (2 \phi) \frac{1}{2}\left(\pi^{xx}_{\nu} -\pi^{yy}_{\nu}\right) -\cos(2 \phi) \pi^{xy}_{\nu}.
\]
These are the central results of this paper, which are applicable to any smooth magnetic textures. 
Notably, in the low density limit and at low temperature, $\tilde P_\nu$ and $\tilde \beta_\nu$ are given by the spin susceptibility perpendicular to $\bm n$ since 
$\pi^{ab}_\nu $ is equivalent to the bare spin susceptibility of the $a_{\bm k \alpha}$ field.   
To make $\tilde \beta_\nu$ finite, anisotropy such as $\pi^{xx}_\nu \neq \pi^{yy}_\nu $ or $\pi^{xy}_\nu \neq 0$ is necessary. 
As is known in the spin susceptibility \cite{Leggett1975}, such anisotropy naturally arises with a triplet pairing. They depend on the relative phase between $\Delta^{\uparrow}_{\bm k} $ and $\Delta^{\downarrow}_{\bm k} $
 as $
\frac{1}{2}\left( \pi_{\nu}^{xx}- \pi_{\nu}^{yy} \right) 
= \lim_{\bm q \rightarrow \bm 0}N^{-1} \sum_{ \bm k} {\rm Re}( \Delta^{*\uparrow }_{\bm k } \Delta^{\downarrow }_{\bm k } ) f_\nu(\bm k, \bm q)$ and 
$
\pi_{\nu}^{xy}
= \lim_{\bm q \rightarrow \bm 0}N^{-1}  \sum_{ \bm k} {\rm Im}( \Delta^{*\uparrow }_{\bm k } \Delta^{\downarrow }_{\bm k } ) f_\nu(\bm k, \bm q),
$
where $f_\nu(\bm k, \bm q)$ is presented in the SM \cite{SM}. Therefore, a triplet pairing can make $\tilde \beta_\nu$ finite and cause the non-adiabatic torque without extrinsic scattering processes. Furthermore, $\tilde P_\nu$ and $\tilde \beta_\nu $ depend on the spatial position through the coupling between the $d$ vector and $\bm n$. This is important for a domain wall dynamics as we see below. 

Now, we demonstrate domain wall dynamics\cite{Tatara2004} induced by the obtained STT.     
We consider the Hamiltonian $H_{\rm tot}=H_{\rm el}+H_{\rm spin}$, where $H_{\rm el }$ is given in Eq.~\eqref{ham2} and  
\[
&H_{\rm spin}=\frac{S^2}{2}\nonumber\\
&\hspace{10pt}\times \sum_i \left(-J(\partial_\nu \bm n(\bm r_i, t))^2 -K n_z(\bm r_i, t)^2 +K_{\perp}n_y(\bm r_i, t)^2 \right).
\]
Here, $J$ is the ferromagnetic exchange coupling, and $K, K_\perp$ are the onsite anisotropies that satisfy $  K_\perp \ll K \ll Ja^{-2} $. We consider a domain wall configuration given by 
$\bm n(\bm r , t)=(\cos \phi_0(t) \sin \theta(x,t), \sin \phi_0(t) \sin \theta(x,t),  \cos \theta(x,t) ) $, where
 $ \cos \theta(x,t)=\tanh \left( \frac{x-X(t)}{\lambda} \right) $, $\lambda =\sqrt{J/K}$, and $X(t)$ is  the domain wall center. 
A schematic figure of a domain wall for $\phi_0 =\pi$ is shown in Fig.~\ref{setup}.
With this configuration, the STT in Eq.~\eqref{pol1} is characterized by 
$\tilde{P}_x =\frac{J_{\rm sd} S}{  n_e a^3}
\left[ \frac{1}{2}\left(\pi^{xx}_{x} +\pi^{yy}_{x}\right)  -\pi^{(1)}_{x} \right]$ and $\tilde {\beta}_x =- \frac{J_{\rm sd} S}{  n_e a^3} \frac{\pi^{(2)}_x }{\tilde P_x} $.
Including the effects of damping, we obtain the equations of motion of $\phi(t)$ and $X(t)$ as  
\[
  \partial_{t} X &=\frac{\varv_c }{(1+\alpha^2) }\left( 
 \tau(\phi_0)j_x  +\alpha  F(\phi_0)j_x+\sin 2 \phi_0 
\right),\label{Xdot}\\
 \partial_{t}  \phi_0& =\frac{-1}{(1+\alpha^2) t_0}\left( 
  \alpha \tau(\phi_0)j_x- F(\phi_0)j_x +\alpha  \sin 2 \phi_0 
\right), \label{phidot}
\]
where $\alpha $ is the Gilbert damping constant, $\varv_c=K_\perp \lambda S/2$, and $t_0 =\lambda/\varv_c$. $\tau(\phi_0)$ and $F(\phi_0)$ denote the coupling to the current via STT.  They read 
\[
\tau(\phi_0)&=
-\frac{a}{2 \varv_c}  \sum_i \frac{\tilde P_x a^3}{2eS} \partial_x n^z, \label{tau}\\ 
F(\phi_0)&= 
\frac{a}{2 \varv_c} \sum_i \frac{\tilde P_x a^3}{2eS} \tilde \beta_x \partial_x n^z. \label{F}
\]
In the above equations of motions, we have assumed that the electron spin density induced by $\dot{\bm n}$  does not change the dynamics qualitatively when the spin size $S$ is large. Also, we did not consider pinning potentials for simplicity. 

In the following, we consider triplet pairing given by $\bm d (\bm{k} ) =(-\sin k_y a, \sin k_x a, \delta \sin k_x a)$. 
Such a pairing can be stabilized by the spin-orbit coupling $\bm g(\bm k) \cdot \bm \sigma$ ($\bm g(\bm k)=-\bm g(-\bm k)$) in a system without inversion symmetry; $\bm d(\bm k) \parallel \bm g(\bm k)$  is energetically favored\cite{Frigeri2004}.  
The Rashba type spin-orbit coupling can stabilize $\bm d (\bm{k} ) =(-\sin (k_y a), \sin (k_x a), 0)$, and in our case, it can originate from the boundary between the superconductor and ferromagnet.  Furthermore, we add $d^z(\bm k) = \delta \sin (k_x a)$, 
which can be attributed to the additional spin-orbit coupling due to the broken mirror symmetry about the $xz$ plane.

In numerical calculations, we set $\mu/t = -1.8, \Delta_0/t =5\times 10^{-2}, J_{\rm sd} S/t = 1$, and $T/t= 5\times 10^{-3}$.  
Using the above $\bm d(\bm k)$, we first show $\tilde P_x$ and $\tilde \beta_x$ in the domain wall configuration for $\phi_0 =\pi/4$ (Fig.~\ref{Pandbeta} (a)-(c)). 
Effective spin polarization $\tilde P_x$ is almost constant in space. 
On the other hand, $\tilde \beta_x$ highly depends on the spatial position. 
Importantly, $\tilde P_x$ ($\tilde \beta_x$)  is symmetric (antisymmetric) under $x \rightarrow -x$ for $\delta=0$, while $\tilde P_x$ and $\tilde \beta_x$ are slightly shifted for $\delta \neq 0$. 
Because of such symmetries, $F(\phi_0)$ (Eq.~\eqref{F}) vanishes for $\delta=0$, where we note $\partial_x n^z$ is an even function of $x$. 
On the other hand, for  $\delta\neq 0$ such symmetries of $\tilde P_x$ and $\tilde \beta_x$ are broken, and hence we obtain finite $F(\phi_0)$.
{Similar arguments apply to other $\phi_0$ values. } 
In the following, we will show the resulting domain wall dynamics. 

\begin{figure}
\includegraphics[width=8.8cm]{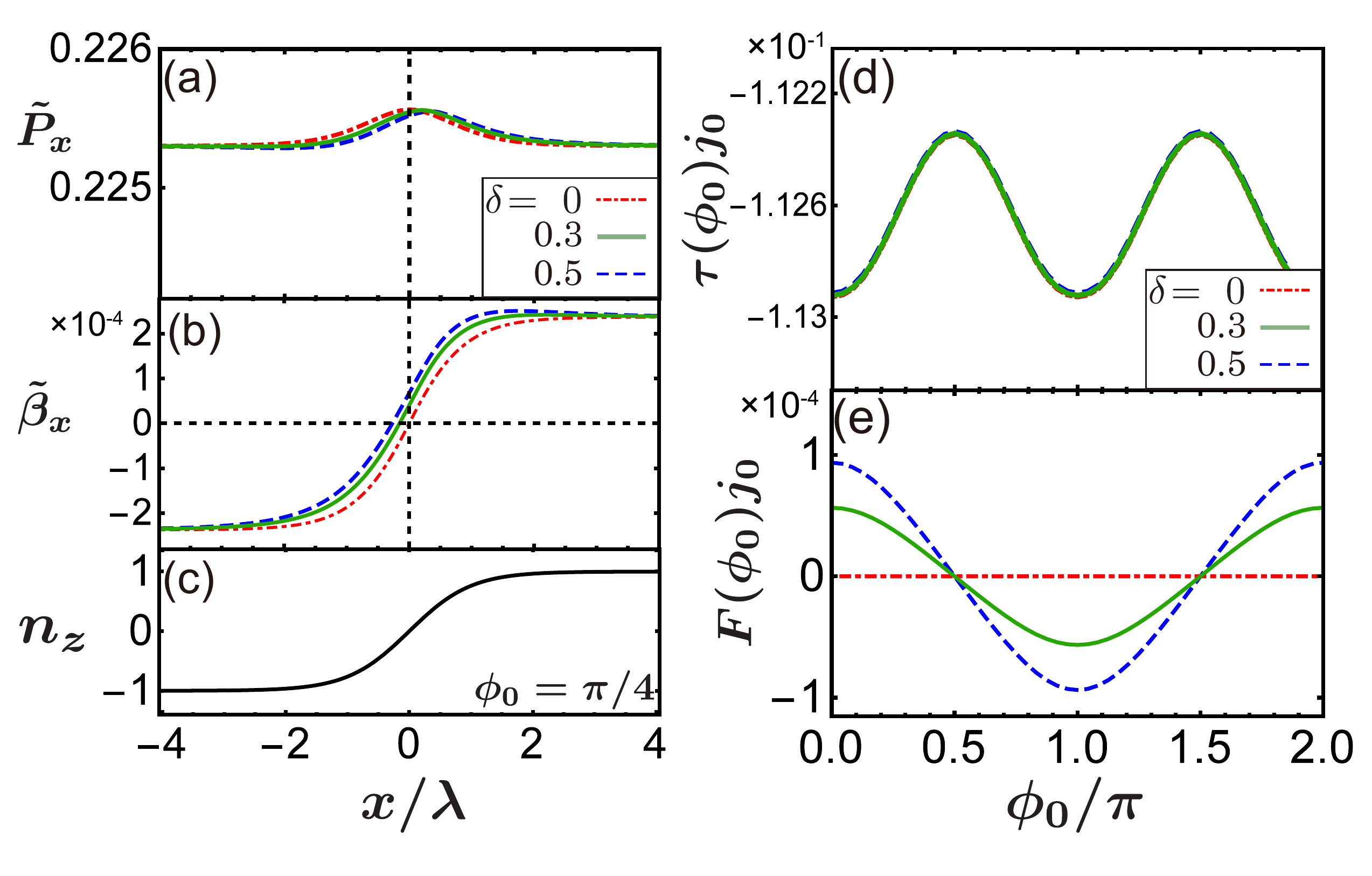}
\caption{(Color online)
(a) $\tilde P_x$ and (b) $\tilde \beta_x$  in a domain wall configuration with $\phi_0 =\pi/4$ as functions of the position $x$ for different $\delta$,  where $d$ vector is $\bm d (\bm{k} ) =(-\sin k_y a, \sin k_x a, \delta \sin k_x a ) $. (c)  The profile of the domain wall. 
(d) $\tau(\phi_0)$  and (e) $F(\phi_0)$ as functions of $\phi_0$ for different  $\delta $, where we define $j_0 =S  e \varv_c a^{-3}$.
\label{Pandbeta}
}
\end{figure}
\begin{figure}
\includegraphics[width=8.5cm]{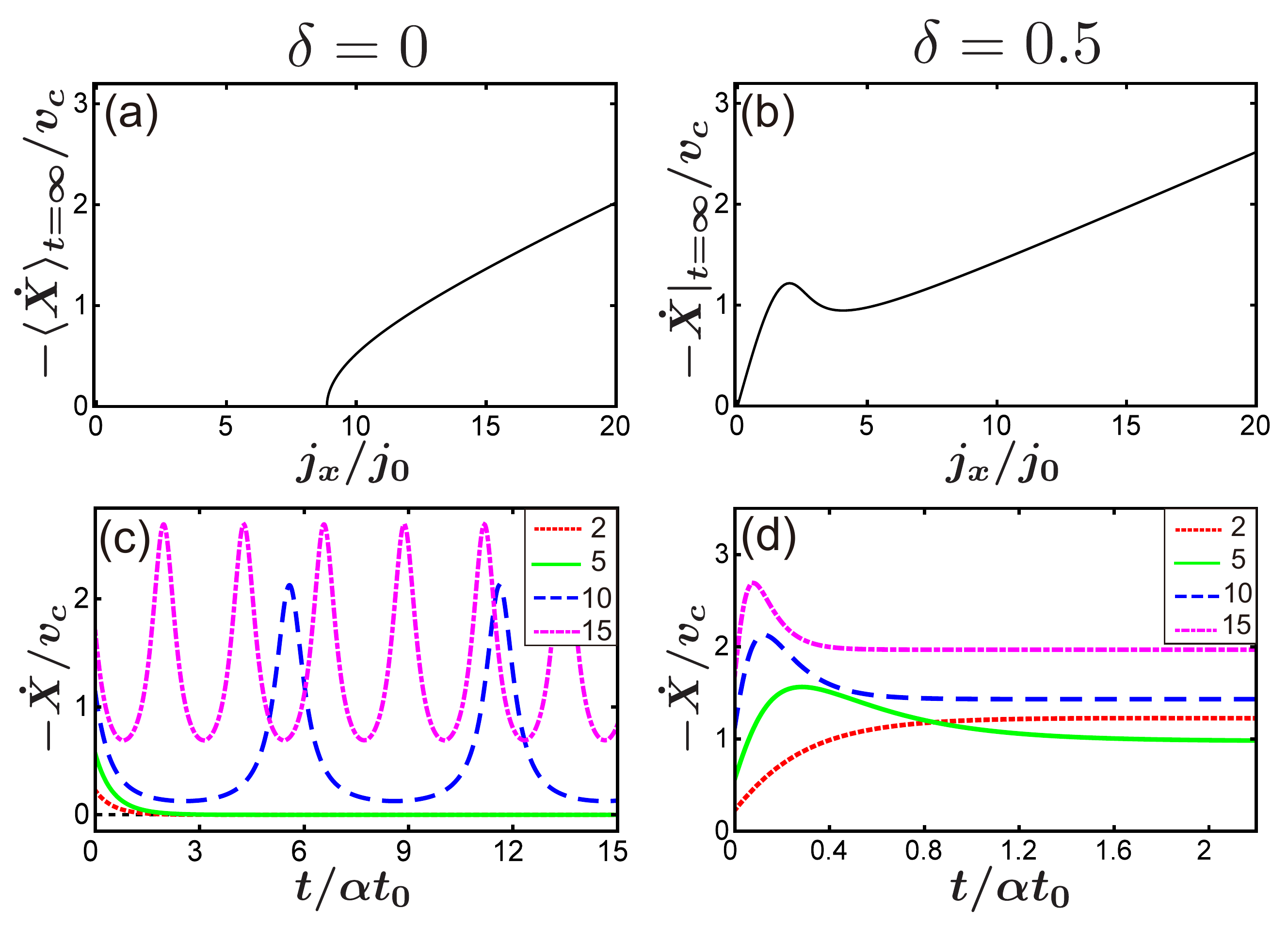}
\caption{ (Color online) Velocity of a domain wall center for  $\delta=0$ in (a) and (c),  and for $\delta=0.5$ in (b) and (d).   
(a)  shows the velocity averaged over the oscillation period after sufficiently long time, and 
 (b) shows the velocity after sufficiently long time. In (c) and (d), the time evolution of the velocity for different $j_x/j_0$ (indicated by different colors) is presented. We have set $\alpha =10^{-4}$, and the initial conditions are $\phi_0 (t=0)=\pi$ and $\dot{X}(t=0)$. 
 }
\label{velocity_f}
\end{figure}

We next solve the equation of motion. For $\delta=0$,   $\tau(\phi_0)$ is well fitted by $\tau(\phi_0) \simeq \tau_0 +\tau_1 \cos2\phi_0$,  and $F(\phi_0)= 0$ as shown in Fig.~\ref{Pandbeta} (d), (e).  
Note that such $\phi_0$  dependence of $\tau(\phi_0)$ arises from the triplet pairing. With this $\tau(\phi_0)$, we have solved Eqs.~\eqref{Xdot} and \eqref{phidot}. 
Fig.~\ref{velocity_f} (a) shows the averaged velocity after sufficiently long time, and there is a threshold current density. It is given by $ j_c= \frac{1}{\sqrt{\tau_0^2 -\tau_1^2} }$, which is obtained from Eqs.~\eqref{Xdot} and \eqref{phidot} with $\tau(\phi_0) = \tau_0 +\tau_1 \cos2\phi_0$ and $F(\phi_0) =0$. As shown in Fig.~\ref{velocity_f} (c), $\dot X$ is zero after sufficiently long time  for $j_x < j_c$ because of the intrinsic pinning due to the anisotropy $K_\perp$. 
With a larger current density ($j_x>j_c$), the domain wall center oscillates with a finite drift velocity. 
The above behavior is similar to a domain wall motion in normal ferromagnetic metals without the non-adiabatic torque.  

We now consider the case with $\delta \neq 0$. As shown in Fig.~\ref{Pandbeta} (d) and (e), $\tau(\phi_0)$ and $F(\phi_0)$  are well fitted by 
\[
\tau(\phi_0) &\simeq \tau_0 +\tau_1 \cos2\phi_0,\label{tp}\\
F(\phi_0)&\simeq F_0 \cos \phi_0. \label{Fp}
\]
Importantly, we have finite $F(\phi_0)$ when $\delta \neq 0$.  As shown in SM \cite{SM}, finite $d^x(\bm k) d^z(\bm k) $ and $ d^y(\bm k) d^z(\bm k)$ are necessary to have finite $F(\phi_0)$.  

This $F(\phi_0)$  changes the domain wall motion drastically. As shown in Fig.~\ref{velocity_f}(b), there is no threshold current density for driving the steady motion, which is similar to a situation in normal metals with the non-adiabatic torque.
An important difference from conventional STT is that a domain wall shows no oscillatory motions ( $\ddot{X}=\dot{\phi} =0$) even for a large current density depending on $\alpha$. 
According to Eqs.~\eqref{phidot}, oscillation occurs for $j> j_{\rm max}= \max_{\phi_0} \left\{ \alpha \sin 2\phi_0 /(\alpha \tau(\phi_0)-F(\phi_0) \right)\}$. 
With the numerically obtained parameters in Eqs.~\eqref{tp} and \eqref{Fp}, $j_{\rm max}$ is infinite when $\alpha $ is small enough ( $\alpha \lesssim 10^{-3}\sim |\Delta_0|^2/t^2 $) because $\alpha \tau(\phi_0)-F(\phi_0)$ can be zero.
In contrast, for conventional STT, $j_{\rm max}$ is always finite since $ \tau (\phi_0)$ and $F(\phi_0) $ do not depend on $\phi_0$, and oscillatory motion always appears for a current density larger than $j_{\rm max}$. 
The absence of oscillatory motion is important for an efficient manipulation of a domain wall. 
Let us estimate the required supercurrent density. For example, in a ferromagnetic nanowire Ni$_{81}$Fe$_{19}$, an experimental value is  $S^2 K_{\perp}\lambda a^{-3} \sim 0.05 $ J/m$^2$\cite{Yamaguchi2006}, and hence $\varv_c \simeq 3\times 10^2$m/s and $j_0 \sim 4 \times 10^{13}$A/m$^2$.  The required current density  to achieve $\dot{X}\simeq 0.4\  \rm{\mu m/s}$ is $j_x \simeq 10^5 $A/m$^2$, which is lower than the critical current density in typical ferromagnetic Josephson junctions\cite{Oboznov2006}. 

To summarize, we have microscopically derived STT induced by triplet supercurrents.  
We showed that spin-triplet pairings give novel types of STT, which can be used for an efficient control of a domain wall.   
The results can be applied to different $d$ vectors and magnetic textures such as a skyrmion, and we expect the possibilities for more interesting aspects of triplet supercurrent-induced STT. 
  
There are several comments and discussions. 
In normal metals, a voltage drop occurs due to the domain wall motion\cite{Berger1986, Volovik1987, Barnes2007, Duine2009} in addition to the resistance of a sample.  
For a superconducting system, while a supercurrent (dc current) is not accompanied by the voltage drop, the motion of a domain wall would also cause time evolution of the phase, and it might result in a finite voltage drop.
In this work, we have assumed such fluctuation is small compared to the overall phase gradient.

In this paper, we did not consider the Abrikosov vortices in a
superconductor, which might be induced by the stray field of the
ferromagnet. These vortices can cause voltage drop due to their dynamics and
a non-uniform current pattern. To suppress the vortices, we can use
junctions with a ferromagnet with a small stray field, e.g.,
Sr$_2$RuO$_4$/permalloy junction with magnetization oriented in-plane. 
We can also use a singlet superconductor with a high critical field \cite{Bergeret2005} such
as niobium, and hence Nb/Ho/permalloy junction is another possible setup.

When the superconducting pairing is proximity-induced, in general, singlet pairing is expected to be mixed. However, the decay length of the singlet proximity effect is much shorter than that of the equal spin triplet pairing for a large exchange coupling. Furthermore, the contribution from singlet pairing to STT is much smaller than that from triplet pairing in the adiabatic regime, which is justified in a smooth magnetic texture. 

{\it Acknowledgements. } We would like to thank M.~S. Anwar, H. Kohno, T. Nomoto, Y. Shiomi, and Y. Yanase for fruitful discussions. 
This work was supported by Grants-in-Aid  for Scientific Research [Grants No.~25220711, No.~17K05517, No.~JP15H05852,  No.~JP16H00988,  (KAKENHI on Innovative Areas from JSPS of Japan ``Topological Materials Science”), and No. JP17H05179 (KAKENHI on Innovative Areas ``Nano Spin Conversion Science”). 

\bibliographystyle{apsrev4-1}

%

\widetext
\begin{center}
\vspace{1cm} 
\textbf{\large Supplemental Material for \\ \vspace{3mm} ``Adiabatic and non-adiabatic spin-torque induced by spin-triplet supercurrent'' }
\end{center}
\setcounter{equation}{0}
\setcounter{figure}{0}
\setcounter{table}{0}
\setcounter{page}{1}
\makeatletter
\renewcommand{\theequation}{S\arabic{equation}}
\renewcommand{\thefigure}{S\arabic{figure}}
\renewcommand{\bibnumfmt}[1]{[S#1]}
\renewcommand{\citenumfont}[1]{S#1}

\section{Derivation of Eq.~(9) }
We start from the action:
\[
S
&= -\frac{1}{2} \sum_{n, \bm k, \bm q} \Psi^\dag_{\bm k+\bm q } (i \epsilon_n)  (\mathcal G_{\rm tot}^{ -1})_{ \bm k+\bm q,\bm k} (i\epsilon_n)
\Psi_{\bm k} (i \epsilon_n),\]
where 
\[
&\Psi_{\bm k}(i \epsilon_n) = 
\begin{pmatrix}
a_{\bm k+\bm Q \uparrow} (i \epsilon_n)\\
a_{\bm k+\bm Q \downarrow} (i \epsilon_n)\\
a^{\dag}_{ -\bm k+\bm Q \uparrow} (-i \epsilon_n)\\
a^{\dag}_{ -\bm k+\bm Q \downarrow} (-i \epsilon_n)\\
\end{pmatrix},\\
&(\mathcal G_{\rm tot}^{-1})_{ \bm k+\bm q,\bm k} (i\epsilon_n)
\nonumber\\
&\hspace{20pt}=\begin{pmatrix}
\left(\left( i { \epsilon_{n}} -\xi_{\bm k+\bm Q} \right)\bm 1 +J_{\rm sd} S \sigma^z\right)\delta_{\bm q, \bm 0} -\varv^\nu_{\bm k+\bm q/2+\bm Q} A_\nu^a(\bm q)\sigma^a
 & -\Delta(\bm{k} )   \delta_{\bm q, \bm 0}\\
-\Delta^\dag(\bm{k} )\delta_{\bm q, \bm 0} &
\left( \left( i \epsilon_{n} +\xi_{-\bm k+\bm Q} \right)\bm 1 -J_{\rm sd} S \sigma^z\right)\delta_{\bm q, \bm 0} 
+\varv^\nu_{-\bm k-\bm q/2+\bm Q}A_\nu^a(\bm q) \sigma^{aT}
 \end{pmatrix},\\
&\hspace{20pt} \simeq
\mathcal G^{-1}_{\bm k} \delta_{\bm q, \bm 0}+ U^{(1)}_{\bm k+\bm q, \bm k}+U^{(2)}_{\bm k+\bm q, \bm k}.
\]
Here we define   
\if0\mathcal G_{\bm k}^{-1}&=\begin{pmatrix}
\left(\left( i { \epsilon_{n}} -\xi_{\bm k} \right)\bm 1 +M \sigma^z\right)
 & -\Delta(\bm{k} ) \\
-\Delta^\dag(\bm{k} )&
\left( \left( i \epsilon_{n} +\xi_{\bm k} \right)\bm 1 -M \sigma^z\right)
 \end{pmatrix} \delta_{\bm q, \bm 0} ,\fi
\[
U^{(1)}_{\bm k+\bm q, \bm k}&=-\varv^\nu_{\bm k}Q_\nu \delta_{\bm q, \bm 0}\mathbb I_{4}
-\varv^\nu_{\bm k+\bm q/2} A_\nu^a(\bm q)
\begin{pmatrix}
\sigma_a & 0\\
0&  \sigma_a^T  \end{pmatrix}, \\
U^{(2)}_{\bm k+ \bm q, \bm k}&=-\frac{\partial \xi_{\bm k+\bm q/2}}{\partial k_\nu\partial k_\eta } Q_\eta A_\nu^a(\bm q)
S^a,
\]
and $\mathcal G_{\bm k}^{-1}$ is defined in the main text. To restore the gauge invariance, we need to include the vector potential $\bm A$ and redefine $\bm{\tilde{Q}}=  \bm Q+\frac{e}{c}\bm A $. Assuming the supercurrent, $\bm j \propto \tilde{\bm Q}$,  is homogeneous,  we can apply the perturbation with respect to $\tilde {\bm Q}$ in the same way.   

The spin density under a superconducting current  is given by $ \tilde s^a_{ \bm q} =\frac{T}{2 N} \frac{1}{2} \sum_{\bm k ,n} {\rm tr}[ S^a \mathcal G_{{\rm tot,}\bm k+\bm q, \bm k}(i \varepsilon_n)]$. We calculate it to the linear order of $Q_\eta$ and $A_\nu^a(\bm q)$, and  obtain 
$\delta \tilde s^a_{\bm q}  =   \pi_{\nu \eta}^{ab} A^b_\nu(\bm q) \frac{ j_\eta}{2en_e } $. Here 
\[
\pi_{\nu \eta}^{ab} =\lim_{\bm q\rightarrow \bm 0}
\frac{-T  }{4N ta^2} \sum_{n, \bm k}  \frac{\partial^2 \xi_{\bm k} }{\partial k_\nu \partial k_\eta}
{\rm Tr} [ S^a {\mathcal G_{\bm k+\bm q}(i {\epsilon_n}) S^{b}   \mathcal G_{\bm k}(i {\epsilon_n})}]+\delta^{ab} L^a_{\nu \eta},\label{chiA}\]
where 
\[
L^x_{\nu \eta}&=L^y_{\nu \eta}=\frac{ 1}{ 2J_{\rm sd} S t a^2} \frac{1}{N} \sum_{\bm k} \varv^\nu_{\bm k}\varv^\eta_{\bm k}\left(
\left.\frac{\partial n_F(\varepsilon)}{\partial \varepsilon}\right|_{\varepsilon=E^\uparrow_{\bm k}}
-\left.\frac{\partial n_F(\varepsilon)}{\partial \varepsilon}\right|_{\varepsilon=E^\downarrow_{\bm k}}
\right),\\
L^z_{\nu \eta}&=\frac{ -1}{ 2 t a^2} \frac{1}{N} \sum_{\bm k} \varv^\nu_{\bm k}\varv^\eta_{\bm k} \sum_{\sigma=\uparrow, \downarrow}\frac{\xi^\sigma_{\bm k}  }{E^{\sigma}_{\bm k}}
\left.\frac{\partial n_F(\varepsilon)}{\partial \varepsilon}\right|_{\varepsilon=E^\sigma_{\bm k}} \left(2 n_F(E_{\bm k}^\sigma)-1 \right) ,
\]
with $E_{\bm k}^\uparrow =\sqrt{(\xi_{\bm k} -J_{\rm sd} S)^2 +|\Delta^{\uparrow}_{\bm k}|^2}$, $E_{\bm k}^\downarrow =\sqrt{(\xi_{\bm k} +J_{\rm sd} S)^2 +|\Delta^{\downarrow}_{\bm k}|^2}$, and $n_F(\varepsilon)=(e^{\varepsilon/T}+1)^{-1}$.
$L^a_{\nu\eta}$ are the contributions from the Fermi surface; they are proportional to the derivative of $n_F(\varepsilon)$ and vanishingly small   
at low temperatures in systems with a full gap or point nodes.
In the main text, we neglect $L^a_{\nu \nu}$ considering a low temperature compared to the superconducting critical temperature. 
We note that for $\Delta(\bm k)=0$, two terms in Eq.~\eqref{chiA} cancel each other and $\pi^{ab}_{\nu \eta} =0$.  

According to Eq.~(14) in the main text,
finite $\frac{1}{2}( \pi^{xx}_{\nu} -\pi^{yy}_{\nu}) 
$ or $\pi^{xy}_{\nu} $ are necessary for $\tilde \beta_\nu \neq 0$, and they depend on the relative phase  between  $\Delta^{\uparrow}_{\bm k} $ and $\Delta^{\downarrow}_{\bm k} $ as  
\[
\frac{1}{2}( \pi^{xx}_{\nu} -\pi^{yy}_{\nu}) 
&=\lim_{\bm q \rightarrow 0} \frac{T}{N ta^2} \sum_{\bm k ,n}\frac{\partial^2 \xi_{\bm k+\bm q/2}}{\partial^2 k_\nu }
\left(
\frac{ {\rm Re} (\Delta_{\bm k+\bm q}^{\downarrow }  \Delta_{\bm k}^{\uparrow *} )  }{ (\epsilon_n^2 +E_{\bm k+\bm q }^{\downarrow2}) (\epsilon_n^2 +E_{\bm k}^{\uparrow2}) }
\right),\\
&=\lim_{\bm q \rightarrow 0} \frac{1}{N }\sum_{\bm k}  {\rm Re} (\Delta_{\bm k}^{\downarrow }  \Delta_{\bm k}^{\uparrow* })  f_\nu(\bm k, \bm q),\\
\pi^{xy}_{\nu}
&=\lim_{\bm q \rightarrow 0} \frac{T}{ Nta^2} \sum_{\bm k ,n}\frac{\partial^2 \xi_{\bm k+\bm q/2}}{\partial^2 k_\nu }
\left(
\frac{ {\rm Im} (\Delta_{\bm k+\bm q}^{\downarrow }  \Delta_{\bm k}^{\uparrow*} )  }{ (\epsilon_n^2 +E_{\bm k+\bm q }^{\downarrow2}) (\epsilon_n^2 +E_{\bm k}^{\uparrow2}) }
\right).\\
&=\lim_{\bm q \rightarrow 0} \frac{1}{N} \sum_{\bm k}  {\rm Im} (\Delta_{\bm k}^{\downarrow }  \Delta_{\bm k}^{\uparrow* })  
f_\nu(\bm k, \bm q),
\]
where 
\[f_\nu (\bm k, \bm q) =\frac{T}{ta^2}\sum_{n}\frac{\partial^2 \xi_{\bm k+\bm q/2}}{\partial^2 k_\nu  }
\frac{1}{ (\epsilon_n^2 +E_{\bm k+\bm q }^{\downarrow2}) (\epsilon_n^2 +E_{\bm k}^{\uparrow2}) }.\] 

\section{ $F(\phi_0)$ for a domain wall}

Let us consider $F(\phi_0)$ in a domain wall configuration when $ d^\nu(\bm k) \in \mathbb R$.
 We have
\if0\[
\pi_x^{(2)}= &\lim_{\bm q\rightarrow \bm 0} \frac{2|\Delta_0|^2}{N} \sum_{\bm k}f_x(\bm k, \bm q) 
\left[\sin \phi_0 d^y(\bm k)d^z(\bm k) -\cos \phi_0 d^x(\bm k)d^z(\bm k)\right]\sin \theta\nonumber\\
&+\lim_{\bm q\rightarrow \bm 0} \frac{2|\Delta_0|^2}{N} \sum_{\bm k}f_x(\bm k, \bm q)  
  \cos\phi_0 \sin \phi_0  \left( -d^{x}(\bm k)^2+d^{y}(\bm k)^2 \right) \cos\theta,
\label{pi2_A}
\]\fi
\[
F(\phi_0)= &\frac{|\Delta_0|^2 a^4}{2e\varv_c S}\sum_i {\partial_x n^z} \lim_{\bm q\rightarrow \bm 0} \frac{1}{N} \sum_{\bm k}f_x(\bm k, \bm q) 
\left[ d^y(\bm k)d^z(\bm k)\sin \phi_0 - d^x(\bm k)d^z(\bm k)\cos \phi_0\right]\sin \theta\nonumber\\
&+\frac{|\Delta_0|^2 a^4}{2e\varv_c S} \sum_i {\partial_x n^z} \lim_{\bm q\rightarrow \bm 0} \frac{1}{N} \sum_{\bm k}f_x(\bm k, \bm q)  
 \left[- \frac{1}{2}\left( d^{x}(\bm k)^2-d^{y}(\bm k)^2 \right)\sin(2 \phi_0)
 + d^{x}(\bm k)d^{y}(\bm k)\cos(2\phi_0)
\right] \cos\theta,
\label{pi2_A}
\]
where we have used Eq.~(5)  in the main text. 

In the following, we show that $F(\phi_0)=0$ for $\bm d(\bm k)=(-\sin (k_y a), \sin (k_x a), 0 )$. 
The first line in Eq.~\eqref{pi2_A} is zero since $d^z(\bm k)=0$.   
We note that in a domain wall configuration, $\theta(x)=\pi- \theta(-x)$ is satisfied.  
{Since $f_x(\bm k, \bm q)$, which depends on $\bm n$ through $|\Delta^{\uparrow}_{\bm k}| =|\Delta^{\downarrow}_{\bm k}| =|\Delta_0|^2|\bm d(\bm k)\times \bm n|$ in $E_{\bm k}^\sigma$,  and  $\partial_x n^z$ are invariant} under the spatial reflection ($\theta \rightarrow \pi-\theta $), the second line in Eq.~\eqref{pi2_A} vanishes after the spatial summation $\sum_i$. 

For $F(\phi_0)\neq 0$, finite $d^x(\bm k)d^z(\bm k)$ or $d^y(\bm k)d^z(\bm k)$ is necessary. In this case, the first line in Eq.~\eqref{pi2_A} is nonzero, and  $|\bm d(\bm k)\times \bm n|$ changes its value under $\theta \rightarrow \pi-\theta $ so that the second term is also nonzero in general.  As we show in the main text, $\bm d(\bm k)=(-\sin (k_y a). \sin (k_x a), \delta \sin (k_x a))$ is one way to satisfy this condition.

\end{document}